\documentclass[journal,letter,twoside]{IEEEtran}
\usepackage{amsmath,amsthm,amsfonts,bm}
\usepackage{setspace}
\usepackage{fancyhdr}

\begin{document}

\title{\huge Reinforcement Learning for Hardware Security:\\ Opportunities, Developments, and Challenges}

\author{Satwik~Patnaik, Vasudev~Gohil, Hao~Guo, and Jeyavijayan~(JV)~Rajendran\\[0.5ex]
Electrical \& Computer Engineering, Texas A\&M University, College Station, Texas, USA\\[0.5ex]
\normalsize{\{satwik.patnaik, gohil.vasudev, guohao2019, jv.rajendran\}@tamu.edu}
}

\IEEEtitleabstractindextext{
\begin{abstract}
Reinforcement learning (RL) is a machine learning paradigm where an autonomous agent learns to make an optimal sequence of decisions by interacting with the underlying environment. 
The promise demonstrated by RL-guided workflows in unraveling electronic design automation problems has encouraged hardware security researchers to utilize autonomous RL agents in solving domain-specific problems.
From the perspective of hardware security, such autonomous agents are appealing as they can generate optimal actions in an unknown adversarial environment. 
On the other hand, the continued globalization of the integrated circuit supply chain has forced chip fabrication to off-shore, untrustworthy entities, leading to increased concerns about the security of the hardware.
Furthermore, the unknown adversarial environment and increasing design complexity make it challenging for defenders to detect subtle modifications made by attackers (a.k.a. hardware Trojans).
In this brief, we outline the development of RL agents in detecting hardware Trojans, one of the most challenging hardware security problems. 
Additionally, we outline potential opportunities and enlist the challenges of applying RL to solve hardware security problems.
\end{abstract}

}

\maketitle

\IEEEdisplaynontitleabstractindextext
\IEEEpeerreviewmaketitle

\renewcommand{\headrulewidth}{0.0pt}
\thispagestyle{fancy}
\lhead{}
\rhead{}
\chead{\copyright~2022 IEEE.
This is the author's version of the work.
The definitive Version of Record will be published in\\ 19th International SoC Conference (ISOCC 2022)}
\cfoot{}

\section{Introduction}
\label{sec:Introduction}

\subsection{Reinforcement Learning}
\label{sec:RL}

Reinforcement learning (RL) enables a computing system (a.k.a. \textit{agent}) to learn using a trial-and-error approach by exploring and exploiting the underlying environment.
The agent interacts with the \textit{environment} and gradually realizes how to take improved \textit{actions} to maximize the total expected \textit{reward}~\cite{sutton_barto_reinforcement}.
Over time, the agent learns to take optimal actions sequentially with limited or no prior knowledge regarding the environment.
Formally, RL solves the underlying problem by modeling it as a Markov decision processes (MDP).
An MDP is defined to be a 5-tuple $(\mathcal{X}, \mathcal{A}, P, R, \gamma)$: $\mathcal{X}$ is the set of states; $\mathcal{A}$ is the set of actions; $P(x_{t+1}|a_{t}, x_{t})$ is the probability that action $a_{t}$ in state $x_{t}$ leads to state $x_{t+1}$; the reward function $r_{t+1}=R(x_{t},a_{t})$ gives reward $r_{t+1}$ after taking action $a_{t}$ in state $x_{t}$; the discount rate $\gamma$, $0\leq\gamma\leq1$, discounts future rewards to their present value.

\subsection{Surge of RL in Solving EDA Problems}

The pronounced capability shown by RL-based algorithms in reducing human effort and providing optimal solutions compared to heuristic and algorithm-driven computer-aided design (CAD) tools have led researchers in the electronic design automation (EDA) community to apply RL to domain-specific problems~\cite{budak2022reinforcement}.
For instance, Google demonstrated their RL-based approach could generate optimal chip floorplans in under six hours compared to human-generated floorplans that take months~\cite{googlenature}.
Researchers have utilized RL for problems ranging from logic synthesis~\cite{drills}, optimization of parameters for placement~\cite{agnesina2020vlsi}, global routing~\cite{routingRL}, sizing of transistors~\cite{analogsizing}, gate-sizing to achieve timing closure~\cite{RL-sizer}, etc.

\subsection{Globalization of IC Supply Chain and Associated Threats}

On the other hand, design companies employ the continual shrinking of technology nodes to develop faster and low-power systems, which necessitates access to advanced technology nodes. 
As the financial implications of commissioning, owning and maintaining a state-of-the-art technology node (e.g., 3nm) are exorbitant~\cite{tsmc3nm}, design houses outsource the fabrication of integrated circuits (ICs) to third-party, off-shore, potentially untrustworthy foundries~\cite{rostami2014primer}.
Outsourcing of fabrication leads to security concerns ranging from piracy of the design intellectual property (IP), unauthorized overproduction of ICs, to insertion of malicious circuits (a.k.a. hardware Trojans (HTs))~\cite{rostami2014primer}.
Security researchers widely recognize the insertion of HTs as a pernicious threat. This is because HTs inserted during fabrication cannot be removed, and the damage incurred by HTs has far-reaching consequences~\cite{yang2016a2}.

\subsection{RL for Hardware Security: Opportunities}

Cybersecurity researchers have used RL agents to develop promising approaches for some security problems, including intrusion detection~\cite{RL_intrusion_detection}, fuzzing~\cite{RL_Fuzzing,RL_Fuzzing_USENIX}, and developing secure cyber-physical systems~\cite{RL_CPS1}.
With the latest advancements in RL algorithms and from the perspective of hardware security, such autonomous RL agents are appealing as they can efficiently navigate high-dimensional search space and generate optimal actions in an unknown adversarial environment.
However, using RL for hardware security problems is in its infancy, and researchers have primarily focused on employing RL for detection of HTs~\cite{pan2020test,pan2021automated,gohilDAC22,chen2022adatest} barring recent works on employing RL for insertion of HTs~\cite{gohilCCS22,sarihi2022hardware}.
\section{Detecting Hardware Trojans using RL}
\label{sec:RL_HT_detection}

Logic-testing and side-channel analysis are the two primary classes of techniques used to detect HTs.

\noindent\textbf{Logic testing}-based techniques apply test patterns and monitor the outputs to measure deviations from the expected, \textit{i.e.,} golden output~\cite{pan2021automated,gohilDAC22}.
These techniques suffer from three limitations: (i)~generating test patterns that activate all possible combinational and sequential triggers are challenging, (ii)~detecting HTs that are devoid of any triggering mechanism (e.g., always-on) or HTs without payloads, and (iii)~requiring improvement in controllability and observability through design modifications, resulting in area and power overheads.

\noindent\textbf{Side channel}-based techniques monitor side-channel information (e.g., power, delay) instead of the output response. 
These techniques do not require the HT to be fully activated or propagate its impact to the primary outputs, rendering it a practical approach over logic testing-based HT detection techniques.
These techniques suffer from two limitations: (i)~multiple golden ICs are required to create the golden signature, and (ii)~the impact of HTs on side-channels can be overshadowed by environmental noise (e.g., process variations).
Next, we summarize efforts by researchers in utilizing RL to detect HTs inserted by an untrustworthy foundry.

\noindent\textbf{Test Generation using RL for Delay-based Side-Channel Analysis.} Pan \emph{et al.}~\cite{pan2020test} proposed an RL-based test generation method for delay-based HT detection.
Unlike existing methods that rely on the delay difference of a few gates, this approach utilizes critical path analysis to generate test vectors that maximizes the side-channel sensitivity.
The authors sub-divide the problem of generating effective test patterns to detect HTs into: (i)~how to find a good initial test for triggering the
HT, and (ii)~how to efficiently generate proper succeeding tests to
switch triggering signals.

\noindent\textbf{TGRL.} Pan \emph{et al.}~\cite{pan2021automated} attempt to solve the issues about scalability and detection accuracy by proposing a logic testing-based approach using a combination of testability analysis and RL.
The authors train the RL model using a stochastic learning scheme that generates test patterns, continuously improving itself to cover as many suspicious nodes as possible.
TGRL drastically reduces the test generation time (6.54$\times$ on average) and detects a vast majority of the Trojans in all benchmarks (96\% on average), a significant improvement (14.5\% on average) compared to state-of-the-art techniques.

\noindent\textbf{DETERRENT.} Gohil \emph{et al.}~\cite{gohilDAC22} aim to find a minimal set of test patterns that can activate all combinations of rare nets.\footnote{The problem of determining a minimal set of test patterns is a variant of the set-cover problem, which is NP-complete.}
This is based on the premise that a single test pattern can simultaneously activate multiple combinations of rare nets.
The authors define \textit{compatible} rare nets if there exists a test pattern that can activate all the rare nets (in a given set) simultaneously and develop an RL agent that generates maximal sets of compatible rare nets~\cite{gohilDAC22}.
DETERRENT achieves two orders of magnitude reduction ($169\times$) in the number of test patterns over TGRL~\cite{pan2021automated} while improving accuracy.

\noindent\textbf{AdaTest.} Chen \emph{et al.}~\cite{chen2022adatest} proposed AdaTest that leverages RL and integrates adaptive sampling to prioritize test samples that provide more information for HT detection. 
Such a process progressively generates test patterns with high `reward' values. 
Although AdaTest demonstrates up to two orders of test generation speedup and two orders of test set size reduction compared to the prior works, they do not showcase any results with regards to other RL-based approaches~\cite{pan2021automated,gohilDAC22}.
\section{Challenges in Applying RL}
\label{sec:challenges_RL}

To use RL effectively for hardware security problems, one needs to identify whether the underlying problem can be mapped to an MDP and whether there exists a notion of a sequential decision-making process in the actions.

\noindent\textbf{Problem Complexity.} The RL problem needs to be formulated so that the training process is efficient. 
The complexity of the training process is directly related to the complexity of the underlying game, which is measured either with the state-space complexity or game tree complexity.\footnote{State-space complexity is defined as the number of legal game positions obtainable from the initial position of the game, while the game tree complexity is defined as the number of leaf nodes in the solution search tree of the initial position of the game~\cite{van2002games}.}

\noindent\textbf{Evaluation Time.} The reward computation time for RL should not be the bottleneck; in other words, the computation time for the reward should be quick and inexpensive. 
As opposed to EDA problems, where at times, the reward can be evaluated/computed by running processes within commercial tools, which can be computationally expensive, one needs to figure out methods to accurately characterize the underlying environment to generate rewards quickly. 
To address the challenge of long evaluation time, off-policy
and offline RL methods can be applied, which do not require real-time interaction with the environment. 
Methods such as offline characterization and pre-computation of data can be performed based on domain-specific hardware security problems.

\noindent\textbf{Generality.} The current RL approaches for HT detection produce test patterns for individual benchmarks using separate agents.
Whether a trained model can be transferred to unseen data remains a challenge.
Designing model architectures that can work on unseen data using principles of transfer learning and/or meta-learning are promising directions.
\section{Conclusion}
\label{sec:conclusion}

In this brief, we summarized the efforts undertaken by hardware security researchers in utilizing RL to address one of the consequential hardware security problems, \textit{i.e.,} the detection of HTs inserted by an untrustworthy foundry. 
Then, we outline some general challenges that need to be solved when applying RL to other hardware security problems. 
These challenges and research directions hopefully would inspire future research for employing RL in hardware security, both in the development of attacks and defenses.

\section*{Acknowledgments}
The work was partially supported by the National Science Foundation (NSF CNS--1822848 and NSF DGE--2039610).

\def\bibfont{\footnotesize}
\bibliographystyle{IEEEtran}

\bibliography{main}

\end{document}